# Introducing the Spatial Conflict Dynamics indicator of political violence

Olivier J. Walther[1], Steven M. Radil[2], David G. Russell[3], and Marie Trémolières[4]

[1]University of Florida, Department of Geography, owalther@ufl.edu

[2]University of Idaho, Department of Geography

[3]Independent Researcher

[4]Organisation for Economic Co-operation and Development, Sahel and West Africa Club

11 May 2020

**Abstract:** Modern armed conflicts have a tendency to cluster together and spread geographically. However, the geography of most conflicts remains under-studied. To fill this gap, this article presents a new indicator that measures two key geographical properties of subnational political violence: the conflict intensity within a region on the one hand, and the spatial distribution of conflict within a region on the other. We demonstrate the indicator in North and West Africa between 1997 to 2019 to show that it can clarify how conflicts can spread from place to place and how the geography of conflict changes over time.

**Acknowledgments:** Funding for the Spatial Conflict Dynamics indicator is provided by the Organisation for Economic Co-operation and Development (OECD) Sahel and West Africa Club (Award AWD05652). This article develops an earlier report drafted by the authors (OECD 2020).

## Introduction

The pivot toward the study of civil war, terrorism, and insurgency since the end of the Cold War has reinvigorated interest in the geography of armed conflicts in the wider conflict studies literature. While the influences of geography on armed conflict is always highly contextual, the literature has coalesced around two key findings regarding the geographic nature of conflict. First, when conflict events are mapped with some precision over time, they have a strong tendency to cluster together in both geographic space and time (Buhaug and Gledistch 2008). Second, and connected to the first, conflicts also tend to spread over space and time (Schutte and Weidmann 2011) with the potential to not remain neatly confined with the territorial geography of states (Gleditsch 2007, Salehyan 2009, Arsenault and Bacon 2015). Taken together, these observations reflect dual truisms about the spatial nature of armed conflict: it is both a geographically uneven and temporally dynamic phenomena.

These findings are the result of several intersecting trends, including the increasing application of sophisticated data visualization and analysis tools of geographic information systems (GIS) and the development of numerous datasets that track the location and timing of political violence (Gleditsch 2020). For example, the well-known ACLED data project now boasts access to a collection of hundreds of thousands of records about discrete acts of violence spanning numerous states across multiple continents and regions (ACLED 2020). And yet, despite the proliferation of both GIS-based tools and GIS-ready data, there has been relatively little in the way of research about the dynamic geography of political violence that aims for more than a descriptive mapping of the presence or absence of violence (Walther et al. 2019).

Of course, such descriptive analyses remain useful in providing basic information about areas and populations affected by conflict or in suggesting issues that may serve as motivating grievances for the belligerents. Nonetheless, taking the investigation of the geography of armed conflict seriously would necessarily move toward a more comprehensive questioning of the space-time dynamics involved and the associated understanding of the spatial and temporal properties at play.

The aim of this article is to address this gap by introducing a new spatial indicator designed for use with political event data. The indicator incorporates two fundamental spatial dimensions of political violence that can vary over time within any given region; we call these conflict intensity and conflict concentration. Conflict intensity, or the amount of violence within a region, can increase or decrease over time according to the military capabilities and political strategies of the actors in a conflict. Conflict concentration, or how violent events are distributed spatially across a region, considers whether events are diffused across a region or, conversely, concentrated on a limited space.

The new Spatial Conflict Dynamics indicator (SCDi) was developed with these ideas in mind. The SCDi can leverage political event data from most space-time event datasets, such as ACLED, and can be used for comparisons across multi-state regions or within a single state context. The SCDi is also adaptable to any size or configuration of underlying areas used for analysis and can utilize any temporal duration desired. In this paper, we illustrate the SCDi with an application to the case of North and West Africa, analyzing over 30,000 discrete events through a 22-year time span and across a 21-state geographical area. We use the SCDi to not just show which regions or places experience the most conflicts but to also demonstrate how the inherent spatiality of conflict can change over time.

The article proceeds as follows. In the next section, we discuss recent efforts to develop new spatially disaggregated event datasets that capture the temporal and spatial dynamics of wars and conflicts and highlight the contribution of the SCDi to this burgeoning literature. We then describe the indicator in detail. Using ACLED data, we next show the indicator's potential in describing the long-term spatial evolution of conflicts in North and West Africa. Our conclusion highlights the key contributions and limitations of our new indicator and avenues for future research.

**Motivation for a new indicator**

The past decade has yielded a greater emphasis on developing geographically referenced political conflict event data (e.g., Schrodt and Yonamine 2013). While past analyses on armed conflict were characterized by state-centric approaches and a general lack of reliable data at the substate level, contemporary studies often build on so-called 'spatially-disaggregated' subnational event data to investigate the onset, duration, and impacts of armed conflicts (Raleigh and Hegre 2009, Buhaug et al. 2009, Tollefsen et al. 2012). In this literature, spatially-disaggregated refers to information about an event that has as much geographic (and temporal) specificity as possible about precisely where (and when) an event has occurred.

Numerous datasets now track subnational expressions of political violence, including the Global Terrorism Database (Lafree and Dugan 2007), the aforementioned ACLED, and the Uppsala Conflict Data Program (Sundberg and Melander 2013). Other more specialized geocoded datasets include the Militarized Interstate Dispute Location or MIDLOC dataset (Braithewaite 2010), the Ethnic Power Relations data (Wucherpfennig et al. 2011), the Geo-PKO dataset on peacekeeping operations (Cil et al. 2019), and the Leadership Changes in Rebel Groups dataset (Lutmar and Terris 2019). What these examples all have in common is a reliance on identifying the location of an event as precisely as possible; events in these databases are always 'geocoded' or assigned geographic coordinates which allows them to be mapped using a geographic information system (GIS).

An advantage of such geocoded data is that they are delinked from any underlying administrative area. In the parlance of spatial analysis and GIS, they are encoded as spatial point data rather than assigned as attributes to two-dimensional geographic areas such as counts of events in a country or region. This means that points can potentially be combined with other spatial organized datasets in different ways or at varying levels for investigation and analysis. For example, the literature often models geocoded event data as a function of different local demographic, political, and economic measures, including factors as diverse as the nature of government, population composition, ethnic divisions, poverty, income, inequality, or number and morale of troops (Hegre et al. 2009, Deny and Walter 2014). Environmental factors such as rainfall and temperature variability, visibility and windy conditions, frequency of droughts, and endowment of natural resources are also commonly used to predict episodes of organized violence around the world (Carter and Veale 2013, Marineau et al. 2018).

This is the underlying approach of efforts like the Violence Early-Warning System (ViEWS) project (Hegre et al. 2019) which assume that an incident of violence is partly a function of the characteristics of the setting where it occurs. Although the role of geography as an influence on human agency is contested in the literature, the incorporation of location-based co-variates in such models recognizes that the resources for political action will always vary geographically and that some relevant influences on agency can be observed and measured in the settings in which a violent event occurs. There remain issues with this approach to incorporating geography into the modeling of political violence, such as assumptions that the relationships between context and action are stationary across a geographic region (see Radil 2019).

Nonetheless, disaggregated geocoded events are now the standard in the literature as they are understood to provide a means to incorporate other location-specific data for modeling purposes.

While geolocated event data are now common for the reasons described above, the geography of a set of events itself is rarely a point of investigation in the literature beyond a basic descriptive mapping of event locations. This means that the inherent spatiality of political event data is largely underexplored even though it is suitable for the application of the techniques of spatial analysis, such as point pattern analysis or PPA. PPA is a set of methods used to detect patterns among a set of locations represented as points in two-dimensional geographic space (e.g., Fotheringham et al. 2007). For example, should a group of violent events be expected to present any form of geographical arrangement, such as clustering or dispersion? Although the tools of PPA are well developed, they are rarely applied to events of political violence (see O'Loughlin and Witmer 2011 and Zammit-Mangion et al. 2013 for important exceptions).

Such investigations of the spatiality of subnational political event data are a necessary first step to consider other important questions, such as how political violence spreads geographically over time. While the literature has established connections between rough terrain (Hendrix 2011), the proximity to porous borders (Walther and Miles 2018), the structure of the road network (Zhukov 2012), and distance from centers of political power and the potential for violence, the overall process of the diffusion of violence remains complex and understudied. For example, Schutte and Weidman (2011) clarified that violence can spread through multiple and different processes, including through both the escalation and relocation of a conflict between belligerents. However, their study was limited to a case of conflict within a single state, which limits the insight that can be applied to conflicts involving groups that span international borders.

In sum, direct investigations of the spatiality of disaggregated and geocoded political conflict event data has lagged behind other applications, such as using event locations to link to other spatial datasets for modeling purposes. Further, while many of subnational factors that purport to explain why violence occurs have been investigated, the processes involved in the spatial diffusion of political violence remain underexplored. These are some of the animating issues behind our efforts and the Spatial Conflict Dynamics indicator (SCDi) is intended to be a first step to help fill this gap.

The SCDi is designed to use disaggregated event data within a time-consistent framework. Because it can be applied to any underlying configuration or regions, including existing administrative units or a systematic overlay, like a uniform grid, it can also complement existing efforts to produce information on political, economic, and environmental variables. However, it is important to note that the SCDi is primarily an *indicator* to explore the spatiality and dynamic evolution of subnational conflict rather than a means of defining a method for regression modeling using disaggregated data (e.g., Tollefsen et al. 2012).

## The Spatial Conflict Dynamics indicator of political violence

The SCDi reflects the central themes about the spatial properties of a point pattern. It is comprised of two complementary metrics that focus on two spatial properties of violence: the relative intensity of conflict within a region or zone (conflict intensity) and the distribution of conflict locations relative to each other (conflict concentration). In the language of PPA, intensity is a 'first-order' property of a point pattern that provides a *local* insight into the distribution of events across a study region. It shows where do the patterns differ. Concentration is a 'second-order' property that provides a *global* insight into how events are expressed. It shows whether the pattern is generally clustered, regular, or dispersed. The methods used for both metrics are described below followed by a discussion of how they are combined by the SCDi.

*Measuring conflict intensity*

A first-order property of a point pattern refers to the intensity of the underlying process that produces events across a study area that can be represented as points at discrete locations (Diggle 2003). Measurements of this property are concerned with the variation in the expression of points across a defined or bounded study region. The intensity of conflict events can be estimated from a point pattern's observed spatial density, which is simply derived from dividing the sum of the number of events in a region by a measurement of the study region's area (such as square kilometers) for a defined time period. This can be applied to the entire study region, yielding a single estimate of the 'global' spatial density.

Spatial density can also be 'localized' by first subdividing the study region into smaller and ideally identically-sized subregions, and then calculating the density for each subregion. Figure 1 illustrates how the local spatial density of violence can vary according to the number of violent events within a hypothetical study region that is divided into six subregions of 10 square km each. This approach is the basis for many area-based methods to investigate point patterns, such as quadrat analysis, a technique that examines whether events are distributed randomly or according to a pre-arranged pattern (e.g., Diggle 2003).

Figure 1. Spatial density can vary across a study region and be estimated locally by dividing the area into smaller subregions.

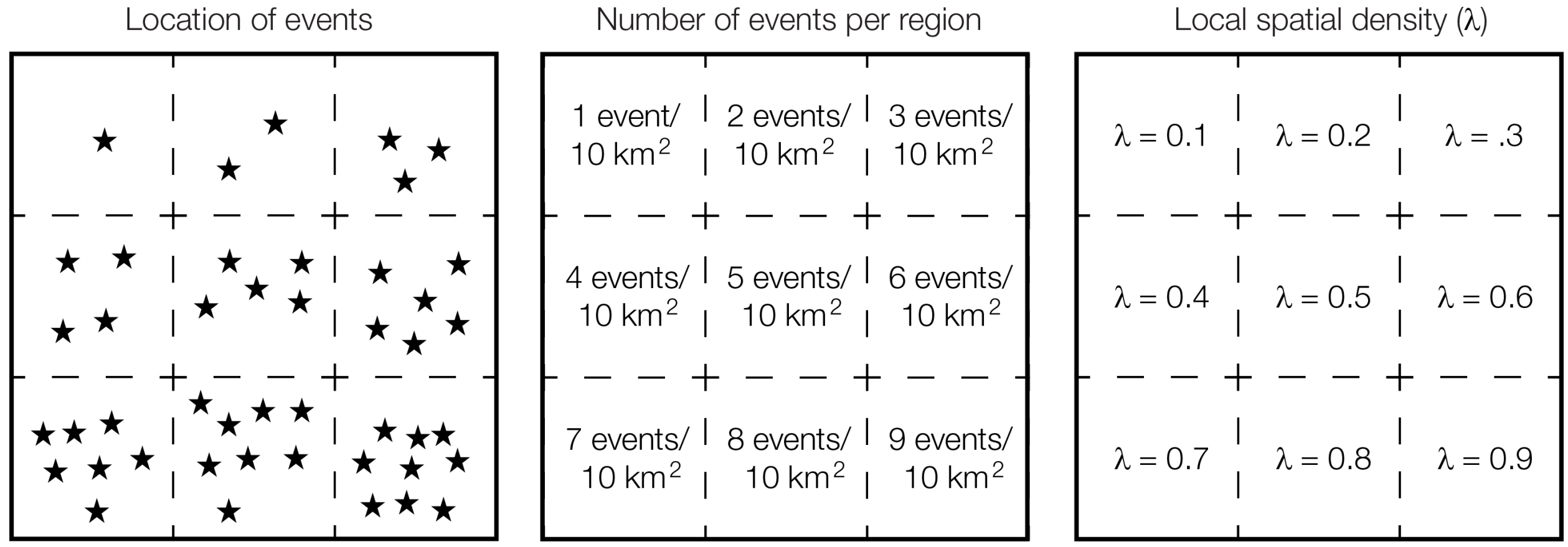

Source: authors

When applied to such smaller subregions, spatial density measures can help identify localized differences in the intensity of conflict across a larger region. These measures can also be used to identify local changes to the intensity of violence over time. From this perspective, spatial density is useful in clarifying trends in violence both over space and time and may be helpful in recognizing the potential implications of an intensification of violence over time in certain subregions. For example, if an increase in the number of violent events takes place adjacent to an international boundary, this may indicate a conflict process is at risk of becoming regionalized or preparing to spread to new regions.

It is important to note that while spatial density is calculated using point data, it is not a measurement of the absolute location of points relative to each other. Instead, it is a measurement of a relative variability of events based on subregional divisions. The size and configuration of these divisions is always the prerogative of the researcher and the imposition of a uniform grid is common. Nonetheless, other choices are possible, including the use of unevenly

sized- and shaped-areas, like political administrative units as the number of events is standardized by the amount of area within the subregion.

This spatial density measurement, which we refer to from here forward as conflict intensity or CI, has a lower bound of 0 when there are no events within a subregion during a given time interval. Because there is no conceptual limit to the number of conflict events that can occur, the CI measure also has no upper bound. Accordingly, as the CI measure increases from 0, it reflects a higher intensity of events within a subregion during a given time interval.

In our study, we use CI scores as one half of the SCDi but they can also be used separately. For example, it is possible to classify a subregion as exhibiting a local CI score that is either higher or lower than expected when compared to some expected CI measurement. To do this, a cutoff value must be established to create a threshold for classification purposes. While this can be done in a number of ways, we illustrate it in our study by first dividing the study region using a uniform grid, then calculating the CI score for every grid cell for every year in the study. We use this set of CI measures to calculate the mean CI score (excluding zeros) over that time span. This resulted in what we called the CI 'generational mean' or the average conflict intensity for a subregion. Therefore, a subregion has a high density if it is greater than the generational mean and a low density otherwise.

*Measuring the concentration of violence*

A second-order property of a point pattern refers to the relative concentration of points to each other across a study area and is concerned with whether points are clustered together or dispersed from each other (Diggle 2003). A variety of measures have been proposed to estimate this property, including those that capture a point pattern's observed average distance between each point and its nearest point. Unlike a localized measure of spatial density then, spatial concentration is a 'global' measure of the propensity for patterning of points relative to each other across an entire study region. In this way, distance-based concentration measures can help identify tendencies in how points are expressed or realized across a larger region.

Most measures of point concentration take departures from a randomly generated pattern (either toward clustering or dispersion) as an open question and the development of statistical methods to evaluate the concentration of point patterns has been a major focus in spatial analysis. However, for most human-caused phenomena that are represented as points, randomness is the exception rather than the rule (e.g., Gatrell et al. 1996). Therefore, we have not emphasized the statistical potential for point randomness in the development of the indicator. Rather, we are concerned with characterizing the amount of either clustering or dispersion in the pattering of conflict events for a given time period.

Among distance-based point-pattern measures, a widely used technique is the average distance to the nearest neighbor (ESRI 2020). With this method, the average distance to a neighboring point is compared against the expected distance under an assumption of a spatially random pattern. If the ratio of the observed average nearest-neighbor distance and the expected distance is less than 1 the pattern is clustered, and if it is greater than 1 the pattern is dispersed. The ratio has a lower bound of 0 with no conceptual upper bound which would represent an extreme geographic clustering of events together in a region (all events at the exact same location). A ratio score of 1 would represent a random pattern of event locations as some would be near each other with others far away but there would be no overall detectible locational pattern. A ratio score more than 1 would represent the relative dispersion of events across the region as they will be further apart from each other than expected.

This is illustrated in the example below in Figure 2. Ten randomly placed events in an area would result in an average nearest neighbor distance of 12 km. Observed-to-expected ratios smaller than one would indicate clustered events while ratios greater than one indicate dispersed events. The distribution of events on the left-hand side of Figure 2, for example, is clustered compared with a random distribution of the same number of events, as shown by its ratio of 0.5. The distribution on the right-hand side is more dispersed, with a ratio of 1.5.

Figure 2. Spatial concentration distinguishes between clustered, random, and dispersed events as measured by the ANN ratio.

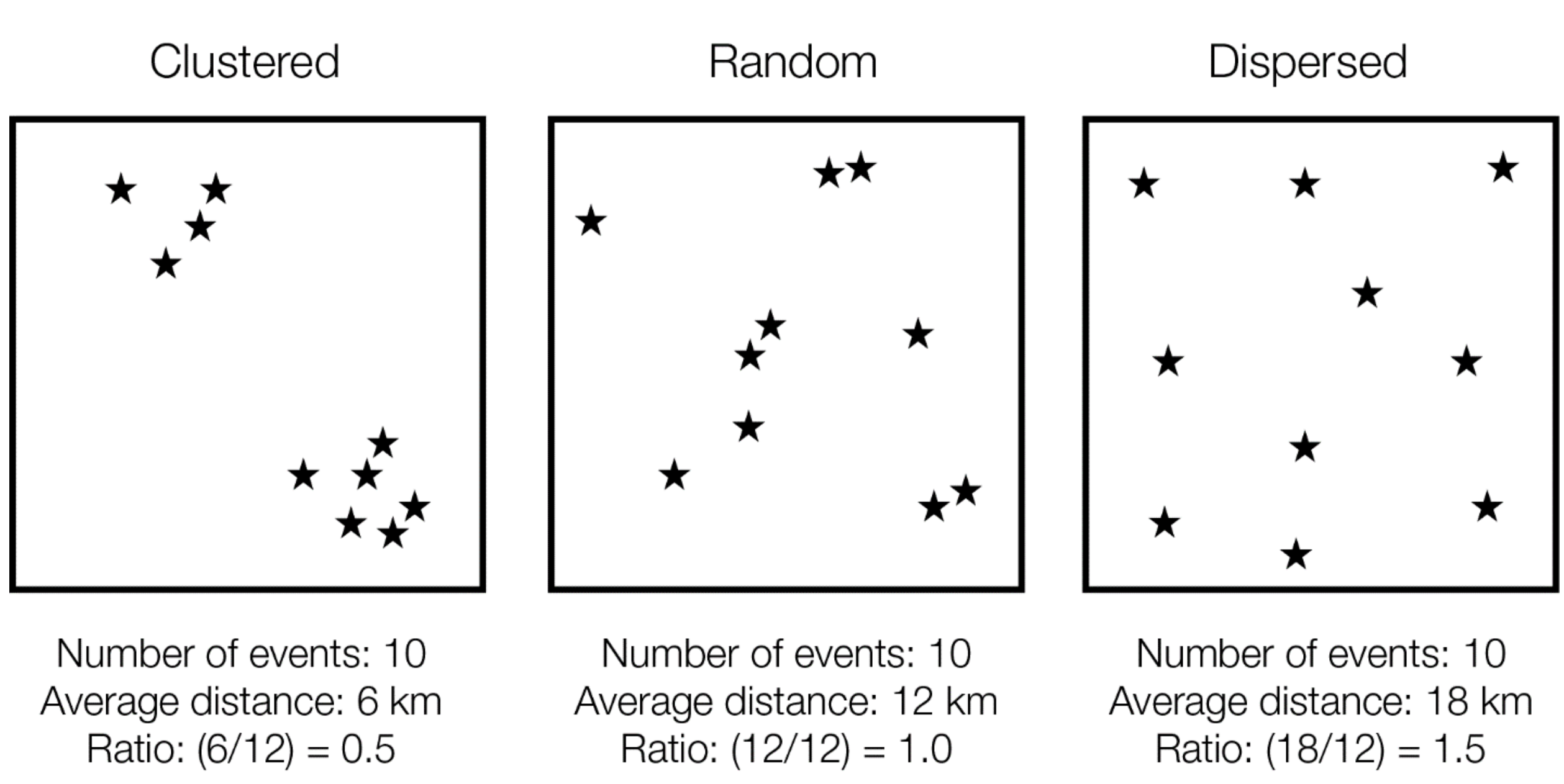

Source: authors

The SCD indicator uses this average nearest neighbor ratio to determine if violent events are clustered or dispersed. While this can be done across an entire study region, it can also be applied using the same division of a study region into subregions as discussed in the previous section. To do this, we first measure the distance between each violent incident in a subregion to its nearest neighbor's location, then average all these nearest neighbor distances. The average nearest neighbor ratio is then used to determine whether the patterns of violent events exhibit clustering or dispersion.

This measurement of conflict concentration (or CC here after) comprises the other half of the SCDi. Similar to conflict intensity, CC can also be used as a stand-alone metric to classify a subregion as exhibiting clustering or dispersion. However, unlike CI, the threshold value to use for classification purposes is already provided by the method. CC scores lower than 1 can be classified as clustered and scores higher than 1 classified as dispersed.

*Four conflict geographies*

By applying the two metrics of conflict intensity and conflict concentration to the same subregion, the SCDi allows the classification of such areas along both a high/low intensity continuum and a clustered/dispersed continuum. This then allows the identification of four different spatial typologies of conflict according to whether violence is more or less intense and dispersed or clustered (Table 1). The first type applies to subregions where there is an above average intensity and a clustered distribution of violent events, suggesting that violence is intense but localized within the subregion. The second type is when a conflict is characterized by a higher than average intensity and a dispersed distribution, indicating that the violence is impacting more locations than with the previous type. The third type applies to subregions where there are fewer violent activities and most of them occur near each other. The fourth type is then when a lower than average intensity and a dispersed distribution of events are combined.

Table 1. Spatial typologies of conflict according to intensity and concentration of events.

| | **High intensity** | **Low intensity** |
|---|---|---|
| **Clustered** | Type 1<br>*More events than mean and closer together than expected* | Type 2<br>*Fewer events than mean and closer together than expected* |
| **Dispersed** | Type 3<br>*More events than mean and further apart than expected* | Type 4<br>*Fewer events than mean and further apart than expected* |

Source: authors.

These spatial typologies are interesting on their own and suggestive of conflict as a political process that occurs in both space and time. For example, when the first type (which combines high intensity and clustering) persists over time, it may indicate that the belligerents are relatively balanced in terms of capabilities with no group able to claim outright control in the subregion. The last type combining low intensity and dispersion may be expected to occur when a conflict is occurring between parties with clear capability imbalances (e.g., hit-and-run style attacks). Such conditions may also be found during the early and late stages of an insurgency and rebellion. In short, these spatial typologies may be connected to certain other issues, such as the capabilities of groups, and to the relative timing of a conflict's 'life-cycle'.

**An application to North and West Africa**

To illustrate the SCDi and the questions it raises, we applied the indicator to a large multi-state region across North and West Africa where, since the early 2000s, a mix of ethnic rebel groups, transnational extremist organizations, and self-defense militias have challenged the legitimacy of several states (OECD 2020). Specifically it was applied to 5 North African countries (Algeria, Morocco, Libya, Tunisia and the Western Sahara) and 17 West African countries (Benin, Burkina Faso, Cameroon, Chad, Côte d'Ivoire, Gambia, Ghana, Guinea, Guinea-Bissau, Liberia, Mali, Mauritania, Niger, Nigeria, Senegal, Sierra Leone, Togo).

While not all countries in the region have experienced significant episodes of armed conflict, many of the region's governments are increasingly confronted with non-state actors that tend to relocate to other countries when confronted by counter-insurgency efforts (Skillicorn et al. 2019). In northern Nigeria, for example, the joint counter-offensive led by Nigeria and its regional allies has led the Jihadist organization Boko Haram to develop its activities in neighboring Niger and Cameroon since the mid-2010s (Thurston 2018). Al Qaeda in the Islamic Maghreb (AQIM) has followed a similar evolution: expelled from northern Algeria by government forces, the Jihadist organization has first relocated to Mali in the mid-2000s, before moving to neighboring Burkina Faso and Niger in recent years (OECD 2020). The geographic spread and opportunistic relocation of such conflicts is amplified by the lack of controls on many African borders that facilitate the circulation of fighters, hostages and weapons (Walther and Miles 2018). In this section, we describe how the SCDi can contribute to understand changes in the evolution of conflict in the region over the last 20 years.

*Data source*

As discussed earlier, the SCD indicator can leverage spatial data from most disaggregated datasets. In this paper, we used event data from the ACLED project, which provides detailed and georeferenced information on actors in armed conflict without imposing a threshold on the number of fatalities recorded for each event (Raleigh et al. 2010). It was applied to the countries listed above from January 1997 to June 2019. We used also calendar years as the temporal interval in our example, which resulting in assessing violence each full year between 1997 and 2018 (22 years).

For our example, we focused on the three main categories of events classified as violent by ACLED: battles, explosions and remote violence, and violence against civilians (Table 2). During the period of observation, these 30,360 events have involved 2551 unique organizations and caused 138,207 fatalities. Demonstrations and non-violent events such as agreements, arrests, disrupted weapons use, headquarters established, looting and non-violent transfer of territory were excluded from the analysis. Apart from excluding the types of events described above, we used ACLED's classifications without modification. Readers are referred to the ACLED codebook (ACLED 2019b) for more detailed information.

Table 2. ACLED violent event types, event counts, and associated fatalities, Jan 1997-Jun 2019.

| Event type | Sub-event type | Events | Fatalities |
|---|---|---|---|
| **Battles** | Armed clash | 12,206 | 59,733 |
| | Government regains territory | 898 | 4748 |
| | Non-state actor overtakes territory | 827 | 4277 |
| **Explosions/Remote violence** | Chemical weapon | 0 | 0 |
| | Air/drone strike | 1487 | 5564 |
| | Suicide bomb | 483 | 4882 |
| | Shelling/artillery/missile attack | 725 | 1176 |
| | Remote explosive/landmine/IED | 2137 | 7605 |
| | Grenade | 53 | 48 |
| **Violence against civilians** | Sexual violence | 119 | 887 |
| | Attack | 9948 | 49,287 |
| | Abduction/forced disappearance | 1477 | 0 |
| **Total** | | **30,360** | **138,207** |

Source: authors based on ACLED data.

*Choosing the right grid for the region*

As previously described, the SCDi can be applied to any group of regions or areas. We adopted the conventional approach in the literature and divided the larger study region into a uniform grid of 6540 subregions (or cells from here forward) of 50 by 50 km. Each cell therefore encompassed the same amount of land area (2,500 square km). In any such endeavor, it is important to explore alternative approaches to defining subregions, which will always necessarily reflect the details of the larger study region (see Figure 3 below). For example, using a much smaller grid of 10 by 10 km across such a vast region would have potentially provided a more granular understanding of violence. However, because political violence is already highly clustered at this scale, more than 90% of the cells this size would have been empty. Of the small proportion of cells that would have had events within them, fewer than half would have had more than one event. The indicator would not have been particularly meaningful with such small numbers of points in each cell. Larger cells, such as a 100 by 100 km grid, are another option but would have aggregated distant events that would not be necessarily linked to one another.

By comparison, a 50 by 50 km grid provided a balanced approach to these issues for our study region. Each cell in the grid was large enough to aggregate a sufficient number of violent events for meaningful analysis while still small enough to provide a localized assessment of political violence across the region. This allows comparing the evolution of conflicts from Dakar to N'Djamena, and from Lomé to Algiers. Using a uniform grid also provided a much more homogeneous representation of violence than existing administrative units, whose size differs enormously across countries and bioclimatic zones. Administrative regions tend to be much bigger in the sparsely populated Sahara than anywhere else, for example, which would greatly affect the density and diffusion of violent events.

Figure 3. Examples of alternative grids and the associated number of events by cells.

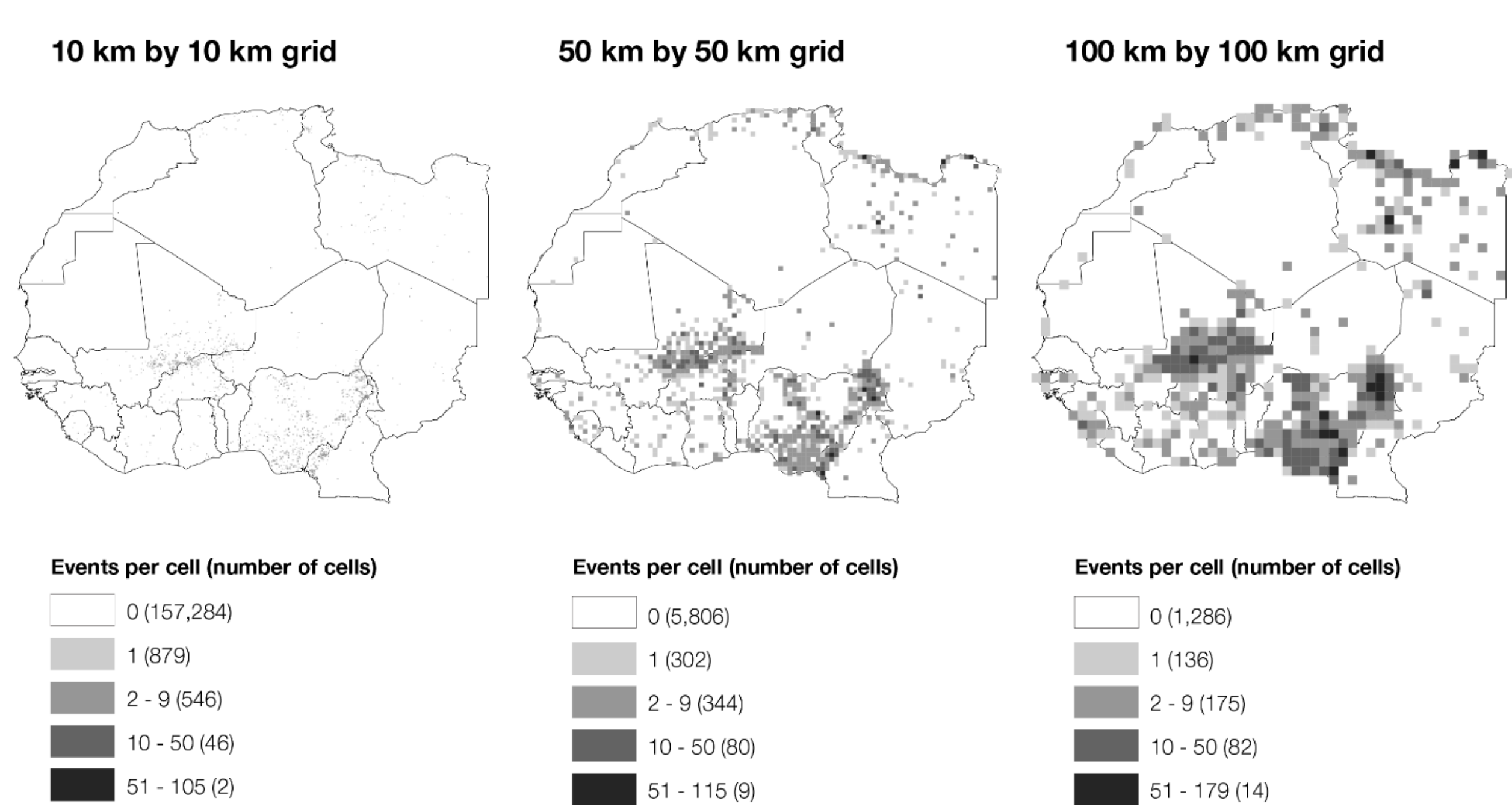

Source: authors based on ACLED data.

The grid was first overlaid onto the study region and the event locations using a GIS. The cells were then used to determine the two separate calculations needed for the indicator: conflict intensity and conflict concentration per cell. Next, we calculated the CI generational mean to use for classification purposes, which was 0.0017 events per square kilometer between 1997 and 2016. Given that the study uses a region of 2,500 square km, a low-density region is then one with four or fewer events in a calendar year. Therefore, when the CI of a subregion was less than 0.0017, it was classified as low-intensity, when it was above that value it was classified as high-intensity.

Calculations were then completed using a Python script developed by the research team. This primarily consisted of two nested loops that first calculated the conflict intensity and then the conflict concentration for each cell, year, and type of event. Because the clustering calculation requires a minimum input of two points, the script also selected only the cells that contained more than one violent event within them. The resulting output for any given year was a grid of subregions with the spatial density and clustering measures for each cell containing two or more events. Pre-processing of the ACLED data was done in R. In the following discussion, we first consider conflict intensity and conflict concentration separately before discussing the full SCDi results.

*Conflict intensity: more subregions are experiencing violence, but intensity has been consistent*

Applying the CI metric to every grid by year shows how the geography of conflict has changed over time in the region and we highlight a few takeaways from examining the CI by itself at the beginning and end of the time range of our study. First, in 1997, most of the cells classified as more intensely violent were within Sierra Leone, with additional small pockets found along the Algerian and Nigerian coasts. However, in 2018, the geography of violence was quite different. Over time, violence had shifted away from most of the aforementioned locations and was found in new places within Mali, Burkina Faso, Niger, Chad, Cameroon, and Libya (Figure 4). This shows the overall geographical dynamism inherent in political violence and why it is important for indicators or measures such as ours to include a temporal as well as a spatial view.

Second, not only has the location of violence shifted over time since 1997, but it has also expanded in scope within the region. In 1997, 85 subregions had two or more events but in 2018 that number had increased to 433. In the context of this example, that represents slightly more than a 500% increase in the total number of subregions with violence. Interestingly, this increase is consistent when considering subregions of both high- and low-intensity. The proportion of

high- and low-intensity violence subregions is nearly the same between the years (59% high-intensity in 1997, 55% in 2018).

Figure 4. Conflict intensity categories in North and West Africa, 1997 and 2018.

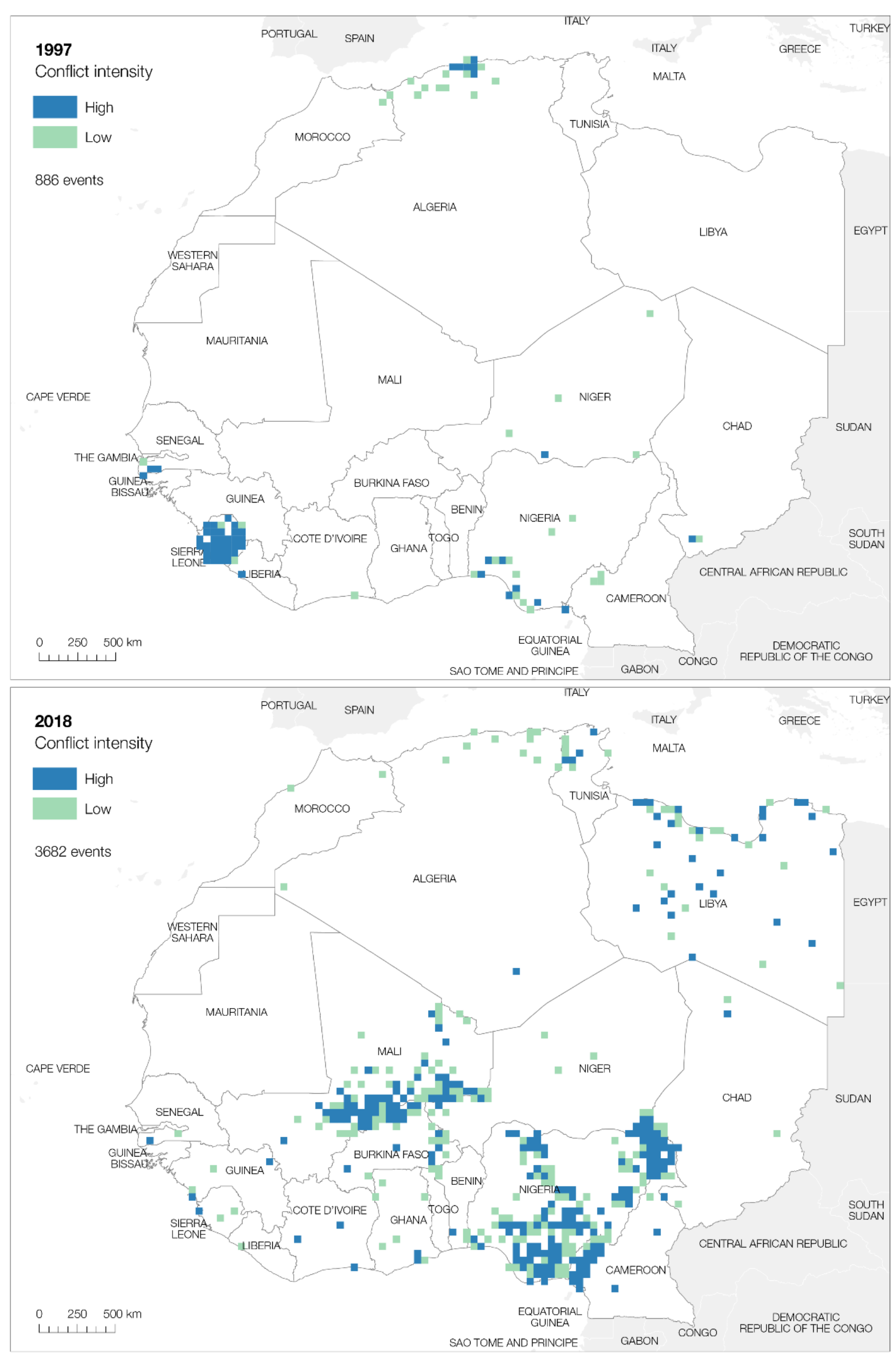

Source: authors based on ACLED data.

Third, while the proportion of low- and high-intensity subregions is relatively unchanged, it is clear that there is a propensity for CI-clustering to occur. Many of the high-intensity subregions form contiguous blocks adjacent to other high-intensity blocks. These groupings tend to be surrounded by similar groupings of low-intensity subregions. In short, the CI shows an inherent spatial tendency in political violence for an intensity gradient to emerge during conflict.

*Conflict concentration: violence patterns have become more dispersed*

Considering conflict concentration by itself also highlights interesting patterns. First, conflict is largely localized within subregions and events are highly likely to occur near one another: the average percentage of subregions classified as clustered (CC < 1) is nearly 91% over the years considered. As a side effect of this geography, the sundry negative impacts of violence are more likely to be felt in the same subregions and populated places repeatedly. In recent years, however, the percentage of subregions that exhibited clustering of violence has dropped from 95% in 2011 to 82% in 2018. This may indicate that violence is become more slightly dispersed. This is a likely consequence of shifting tactics, including the marked increase in the numbers of attacks against civilians as these types of events may be less likely to occur in similar locations over time.

Second, subregions with dispersed violence locations are a particular cause for concern because they may be evidence of the spread of conflict to a new area from a neighboring region. Conversely, a dispersed pattern may be evidence that a conflict is weakening or that one party is dominant in a region, as fewer violent events occur in nearby locations. In other words, a dispersed pattern can identify regions where a transition is underway in either direction. Alarmingly, the percentage of subregions in 2018 with dispersed events (CC > 1) is nearly 17%, the highest percentage of the all the years in the study. This is 7% higher than the historical baseline between 1997 and 2016 and may be a sign of the incorporation of new places into the already intensely localized geography of conflicts. If correct, this would result in a negative feedback cycle by exposing more locations with the region to the effects of violence.

Third, the relative locations of clustered and dispersed pattern regions have also shifted over time. For example, all of the major conflict areas (Figure 5) across the study region include clustered regions in 2018. Over time, dispersed patterns have mostly been associated with conflict zones in Nigeria. The most populous country in Africa, Nigeria is also the country that has, by far, the largest number of violent events (9,017) and fatalities (67,512). Nigeria is home to three major sources of continuous violence that explain the unusual intensity and geographical extent of conflicts in the country: the Boko Haram insurgency in the northeast, communal violence in the Middle Belt, and the Niger Delta insurgency in the south. Taken together, these conflicts account for 30% of the violent events and half of the victims recorded in North and West Africa since 1997.

Figure 5. Conflict concentration categories in North and West Africa, 1997 and 2018.

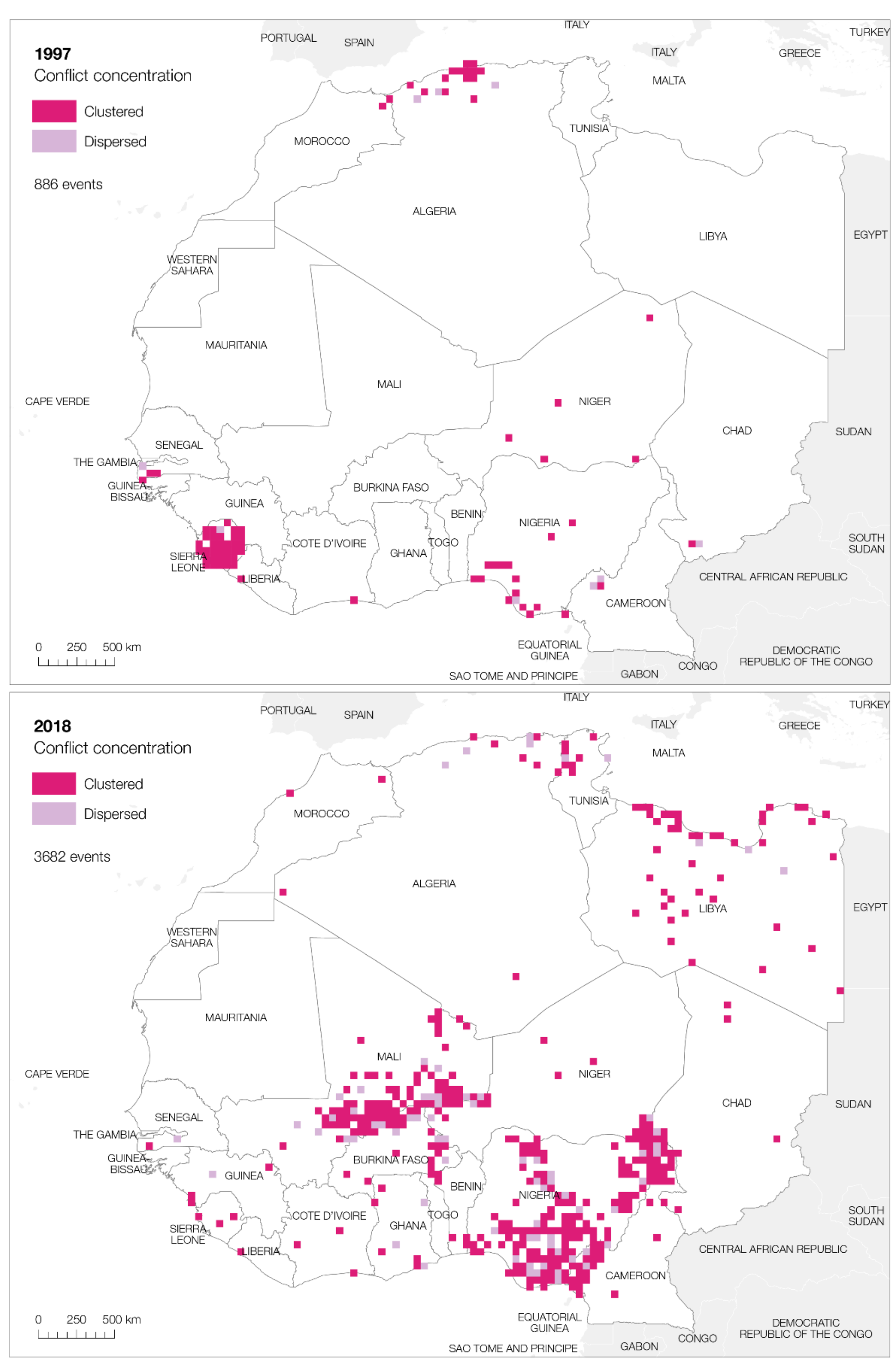

Source: authors based on ACLED data.

*The dynamic geography of political violence*

Since the SCDi combines both measures to identify the four spatial typologies discussed previously, these combined classifications can be used to consider how the geography of conflict has evolved in the region over time. In particular, the SDCi results show that that the last 10 years have been marked by an increase in all types of conflicts in North and West Africa.

The number of subregions experiencing a high-intensity and clustering (type 1) has increased significantly faster over time than the other spatial types. Historically, this type of conflict has been widespread in the region but its proportion to the other types has continuously increased since the mid-2000s, from 38% in 2008 to more than half ten years later (Figure 5). Nowadays, these subregions often form the core of large epicenters of violence, as in central Mali, northern Burkina Faso, around Lake Chad, in the Middle Belt and the Niger Delta in Nigeria, and in Libya (Figure 6).

Subregions in which conflicts are characterized by a high-intensity of dispersed events (type 2), are fortunately quite rare in the region. They concern only 3% of the cells, a proportion that has not changed much over the last 20 years. Several time periods with no cells of dispersed-high density are recorded, indicating that this combination is a rather unusual occurrence in the region. However, the locations where this type has been observed are also where conflict has been the most entrenched in the last decade, including the Inner Niger Delta in Mali, southern Nigeria, the Liptako-Gourma between Niger, Mali and Burkina Faso, and the border region between Nigeria and Cameroon.

Figure 6. Time series of conflict subregions by SCDi categories, 1997-2018.

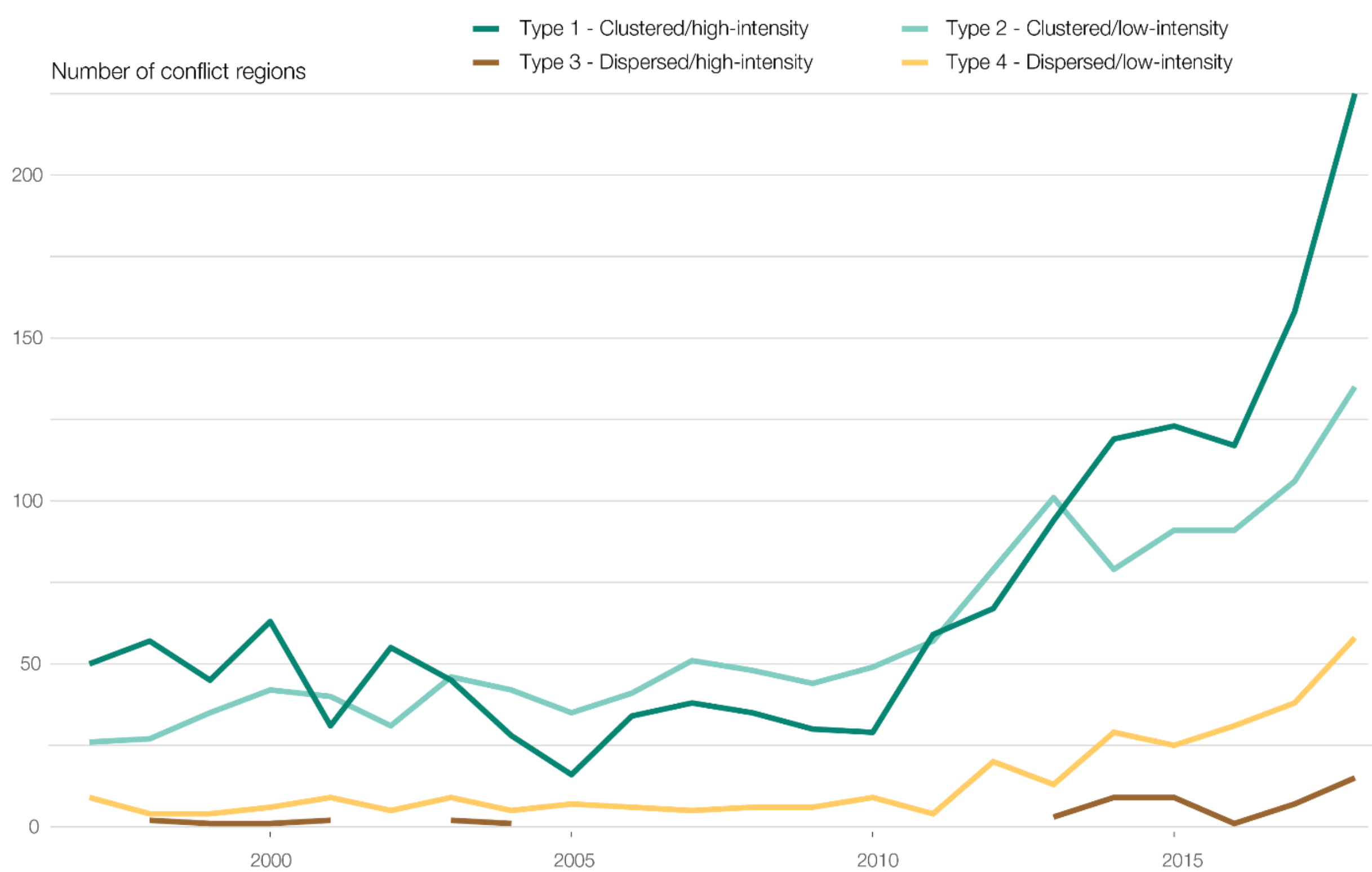

Source: authors based on ACLED data.

The number of subregions in which political violence is both clustered and of low-intensity (type 3) has experienced strong growth since 2010. This type concerns a third (31%) of the cells of the wider region. However, these conflict subregions are still less well-represented than during mid-2000s, when they accounted for half of the four categories. They are often found on the periphery of more intense conflict zones, such as on the outskirts of major cities in Libya.

Subregions that experience dispersed and low-intensity events (type 4) are also rare. This type is found in only 13% of the cells overall but the percentage of those subregions has doubled over the last 10 years. These subregions are located at the periphery of the major war zones or in some countries with fewer violent events, such as Ghana, Guinea or Algeria.

Figure 7. Spatial Conflict Dynamics indicator categories in North and West Africa, 2018.

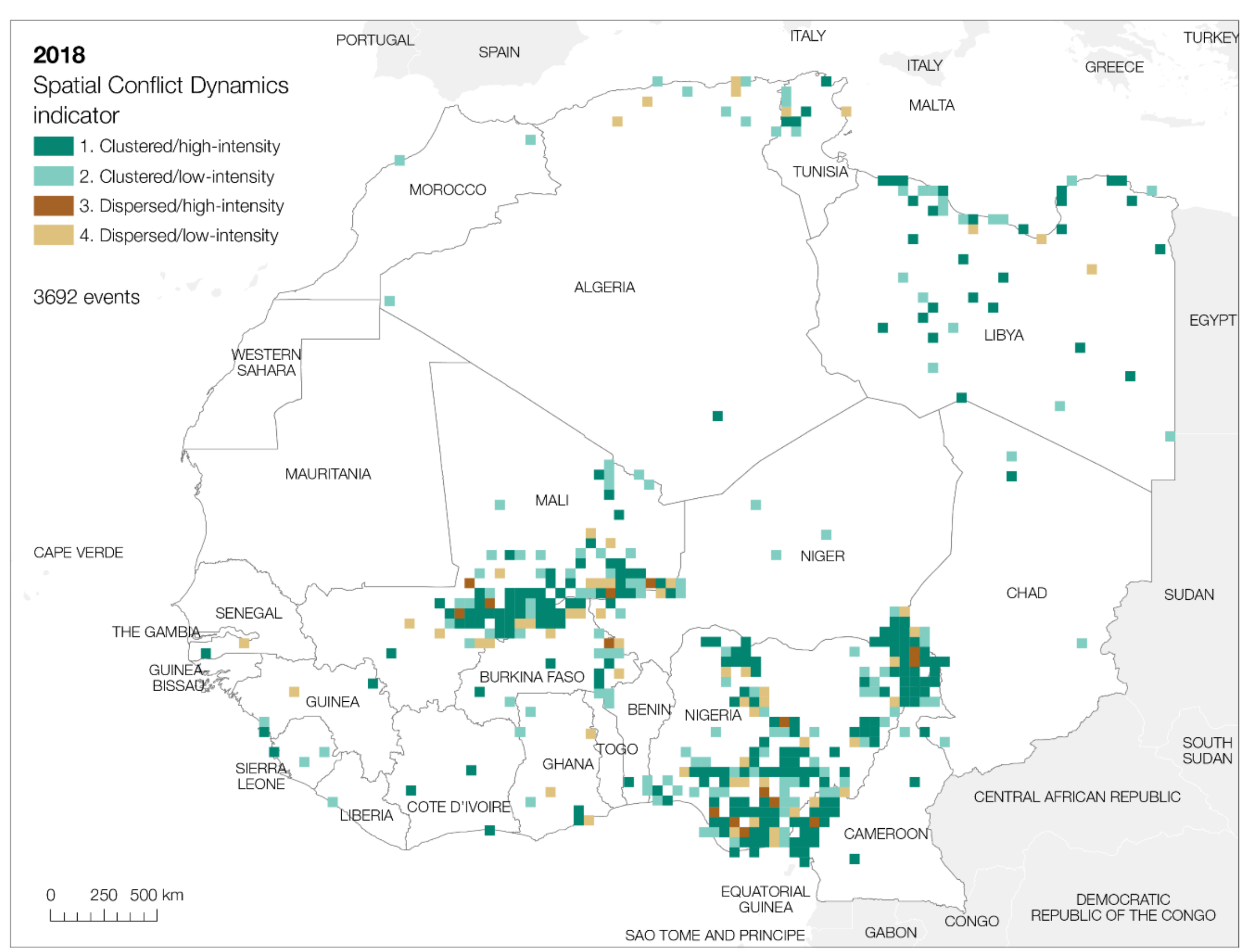

Source: ACLED. Calculations and cartography by the authors.

The changing composition of spatial typologies over time is suggestive of other aspects of conflict geography that remain understudied in the literature. For example, how often does a subregion experience a sudden outbreak of intense and clustered violence? Do conflicts tend to begin with one type and transition to others? And what is the propensity for types to change to something else over time? The SCDi allows the exploration of the dynamics associated with the typologies and such changes are clearly observable when we visualize the year-to-year interplay between the various spatial typologies identified by the indicator. For example, a subregion might have a SCDi classification as clustered/high-intensity in 2003, a score of clustered/low intensity in 2004, and no conflict in 2005.

In Figure 8 below, we tallied these annual shifts across the entire 22 years of the study into an alluvial chart to show how often changes from one type to another occur. Subregions with no or just one conflict event in a given year are excluded in the figure; the remaining subregions are then used to highlight how conflict geography has evolved over time. There are several takeaways from this type of presentation.

Figure 8. Yearly shifts in Spatial Conflict Dynamics indicator categories, 1997 to 2018.

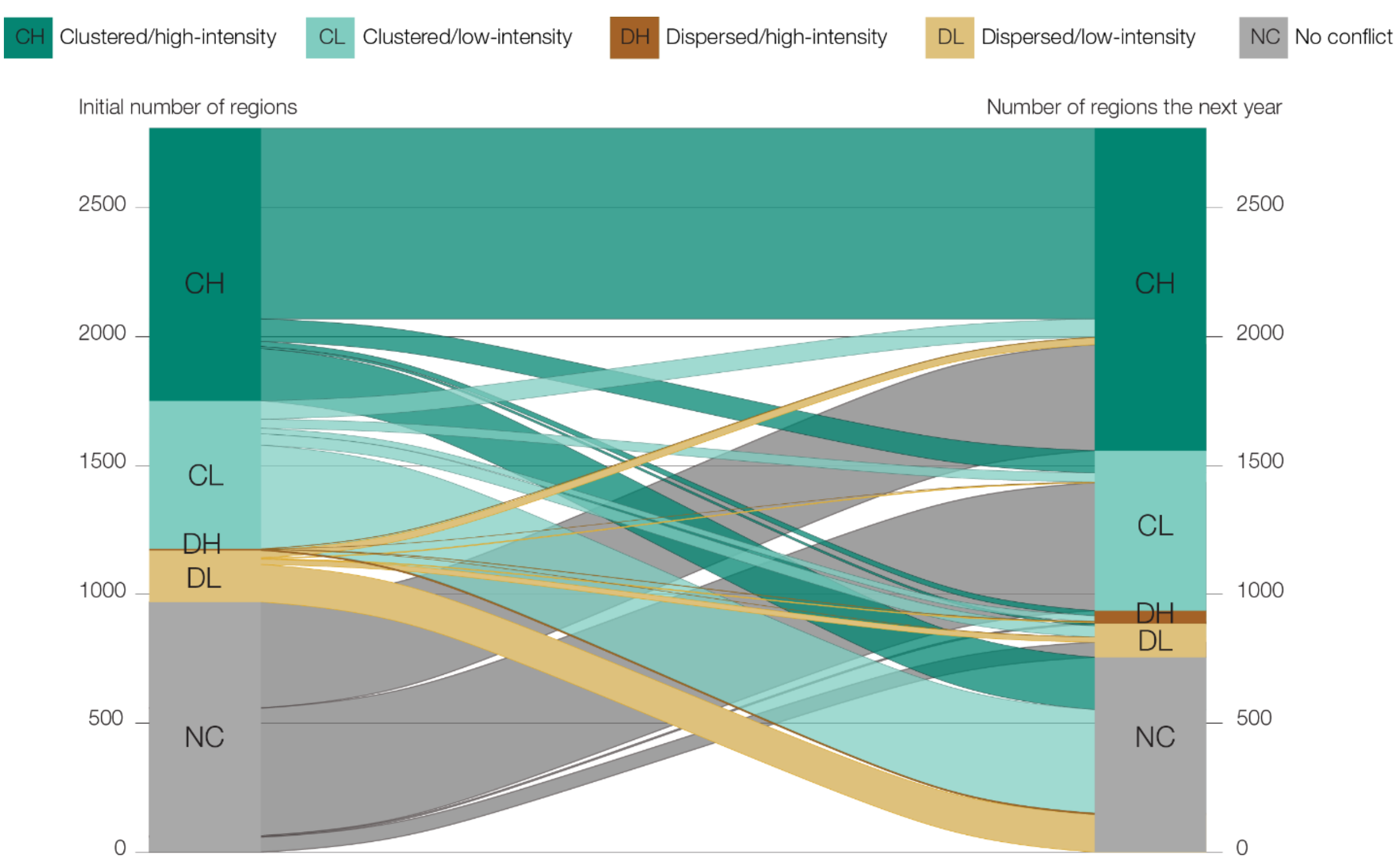

Source: authors based on ACLED data. Note: the figure shows how many regions have experienced a change in conflict from one year to another between 1997 and 2019. Each region

is classified according to its initial and final intensity and concentration of violent events. The figure excludes regions that never receive an SCDi score over any year in the study.

First, the SCDi shows that subregions where violence is both clustered and intense (type 1) are likely to remain unchanged year-to-year. The most common outcome is for a clustered/high-intensity region in one year to remain unchanged in the following year, an outcome which occurs 70% of the time. This reflects the potential for a localized conflict to continue over multiple years, a common occurrence in the various conflict zones in Nigeria for instance.

However, the second most common outcome for a clustered/high-intensity region is to have no conflict the following year (19%), meaning that even intense localized conflicts can come to a rather abrupt end. The third most common outcome for this type is a shift is to clustered/low-intensity (8%), which may be interpreted as a transition state along the way to conflict de-escalation. For example, many of the major cities in Libya (Tripoli, Benghazi, Misratah, Al Khums) continue to have higher than average numbers of events year-to-year. Between 2017 and 2018, the event locations within those conflict zones remained clustered together within the city's extent rather than occurring at dispersed locations around the countryside.

Second, subregions where violence is clustered but of low intensity (type 3) are likely to change for the better. The most common changes for clustered/low-intensity subregions are either to have no conflict (70% of the time) or to become clustered/high-intensity (12%) the following year. This reinforces the idea that clustered/low-intensity subregions may represent a state of transition and are therefore unlikely to persist year-to-year, which only happens 6% of the time. For example, several conflict regions in Yobe and Gombe states in northeast Nigeria changed from low intensity but clustered conflict locations in 2017 to no conflict the following year.

Third, subregions where violence is dispersed and intense (type 2) are also likely to improve. These are the rarest occurring outcomes in the study, a positive sign as they may represent both an intensifying and spreading conflict. Hopefully, the most common outcome is for such subregions to have no conflict the following year (71%) which points to their inherent instability. The other most frequent outcomes are to either remain unchanged or to shift to clustered/high-intensity (each outcome happens 14% of the time). For example, conflict around Biu in Nigeria's Borno state changed from dispersed but high intensity to no conflict between 2017 and 2018. Similar examples over that same time period were found in rural areas of Nigeria's Benue and Edo states and in Chad around Moundou in Logone Occidental.

Fourth, subregions where violence is both dispersed and of low intensity (type 4) are likely to have no conflict the following year. As with the clustered/low-intensity and dispersed/high-intensity categories, the most common outcome for dispersed/low-intensity subregions is to experience no conflict the following year (73%). 14% of the time, these subregions became clustered/high-intensity; 10% of the time, they remain unchanged. For example, between 2017 and 2018, numerous examples of conflict zones changing from dispersed and low intensity conflict to no conflict were found in Nigeria (Bayelsa, Benue, Borno, Edo, and Imo states), in Mali (Timbuktu region), and Libya (near Benghazi).

Lastly, subregions that are not affected by conflicts are likely to remain peaceful but, when a change happens, it is likely to be a localized low intensity conflict (type 3). Most subregions never experienced conflict over the study's time range. However, subregions that did shift from no conflict to one of the other SCDi categories are troubling as this resents a sudden introduction (or perhaps re-introduction) of conflict into a previously peaceful region. When conflict does emerge in a region where it was previously absent, it is most often as clustered/low-intensity (51%). However, it also commonly jumps directly to clustered/high-intensity (42%). This may reflect the worst possible outcome given the persistence of that category over time. For example, several areas in the central part of Mali's Mopti region that had no conflict events in either 2016 or 2017 were classified as clustered and high intensity in 2018. When conflict emerged, it erupted into the category that has the most persistence over time.

**Conclusion**

The Spatial Conflict Dynamics indicator (SCDi) introduced in this article makes it possible to take into account the great diversity of current conflicts in the world. By combining two important measures of the geography of conflict (intensity and concentration), the indicator shows promise to clarify no just where conflict occurs but the different forms it can take when it happens. In North and West Africa, where the indicator was applied for the first time, it also revealed several important elements about long-term political violence in the region that can inform analysts, policy officials and other stakeholders interested in the advancement of peace. Below we discuss what we believe to be some of the important broader implications of our efforts.

First, using the SCDi to track the long-term evolution of the geography of violence highlights that the location of violence is highly dynamic over time. Most of the major conflict areas of the 1990s are peaceful today and much of the current violence is observed in states that were considered stable 15 years ago. This shifting and relocating of political instability, including across international boundaries, should encourage more research on how this diffusion process works even as disaggregated data collection efforts continue to track the locations of violence at fine-grained scales. It also highlights how a regional approach can be helpful; a focus on an individual country or even on a smaller set of states would have missed this essential character of political violence and perhaps failed to detect the direction and implications of such shifts when they occur. For those reasons, the SCDi can be used as a tool to monitor the larger patterns of violence regionally, keeping a watchful eye on the conditions that may be involved with the transitions from peace to violence and vice-versa.

Second, the SCDi illustrates how violence operates geographically over time. The recognition that violence can be differently concentrated while also varying by intensity provides a more nuanced understanding that can shape both governance strategies to deal with the circumstance and affect the relative efficacy of those efforts. For example, a highly clustered expression of violence will result in a different form of impacts on people and places than will a highly diffuse expression. Accordingly, relief and violence suppression efforts will necessarily need to take on a different character to address each. Similarly, understanding the relative intensity of violence over time in the same place or area is an important metric of human security that stakeholders can use to assess the effectives of their response to violence. That the SCDi addresses both aspects of the geography of violence speaks to the potential utility of it as a response tool when there is an outbreak of violence.

Third, the SCDi showed how the spatial dynamics of conflict have played out in aggregate over time. By tracking the SCDi categories in the same subregions over time, we are able to assemble a first glimpse at how conflict has evolved over time within the region. Of course, we recognize that an application of this approach to a different region and/or across a different time period may have yielded different results about the relationships between categories and more work is needed to establish how contingent or contextual these findings may be. Helpfully, the SCDi is flexible and adaptable to other geographical and historical contexts, including the configuration of the subregions used for the analysis. So long as the trends toward the development of disaggregated conflict data continue, the SCDi offers an opportunity to build data-driven profiles about the evolution of conflict within different regions and different historical eras.

The SCDi is not intended as a substitute for grounded qualitative knowledge about conflict in specific places and we recognize that no measure or indicator will be able to perfectly capture all the complexities of the geography of armed conflict. However, we see opportunities to significantly improve upon the initial version of the SCDi. For example, the threshold value we used for the low/high intensity categories was based on data from all subregions across 20 years. This could be refined by utilizing a 'moving window' method, both in space and time, to establish subregion- and year-specific thresholds. This in turn could provide a more nuanced and contextual interpretation of what counts as high- and low-intensity conflict based on the details or both where and when conflict is occurring. The intensity calculation could also be refined through weighting methods to capture variability in the spatial distribution of population or any other factor that may be associated with the production of political violence.

Future efforts to improve and refine the indicator will work to address these concerns. Other planned initiatives will focus on developing an open-source version to allow researchers to easily apply the indicator to other cases and to encourage other innovations not identified here. Ten years after the ACLED database was introduced to the scientific and policy communities, we view the SCDi as an additional tool that can capitalize on the increasing availability of geolocated data to provide a more spatial understanding of conflict around the world.